\documentclass[12pt]{article}
\usepackage[T1]{fontenc}
\usepackage{cite}
\usepackage{mathrsfs}
\usepackage{amsmath}
\usepackage{amssymb}
\usepackage[dvipdfm]{graphicx}
\usepackage{tabularx}
\usepackage{bm}

\newenvironment{minilinespace}{\baselineskip = 8mm}{}
\begin{document}
\begin{titlepage}

\begin{flushright}
{
	KUNS-2131,CAS-KITPC-021
}
\end{flushright}
\vspace{1cm}

\begin{minilinespace}
\begin{center}
	{\large
		{\bf  Stability of Five-dimensional Myers-Perry Black Holes\\
with Equal Angular Momenta
		}
	}
\end{center}
\end{minilinespace}
\vspace{1cm}

\begin{center}
Keiju Murata$^{a,1}$, Jiro Soda$^{a,b,2}$\\
\vspace{.5cm}
{\small {\textit{$^{a}$
Department of Physics, Kyoto University, Kyoto 606-8501, Japan\\
$^{b}$Kavli Institute for Theoretical Physics,
Zhong Guan Cun East Streat 55, Beijing 100080, P.R. China
}}
}
\\
\vspace*{1.0cm}

{\small
{\tt{
\noindent
$^{1}$ murata@tap.scphys.kyoto-u.ac.jp
\\
$^{2}$ jiro@tap.scphys.kyoto-u.ac.jp
}}
}
\end{center}

\vspace*{1.0cm}



\begin{abstract}
 We study the stability of five-dimensional Myers-Perry black holes
 with equal angular momenta which  have an enlarged 
 symmetry, $U(2)$. Using this symmetry, we derive master
 equations for a part of metric perturbations
 which are relevant to the stability. Based on the master equations,
 we prove the stability of Myers-Perry black holes under these perturbations. 
 Our result gives a strong evidence for the stability of
 Myers-Perry black holes with equal angular momenta. 
\end{abstract}

\end{titlepage}

\section{Introduction}
Recently, higher dimensional black holes  have attracted much interest~\cite{Emparan:2008eg}. 
In addition to Myers-Perry black holes~\cite{Myers:1986un}, which have been found 
as the higher dimensional generalization of Kerr black holes, 
there are many exotic black holes 
in higher dimensions~\cite{Horowitz:1991cd,Emparan:2001wn,
Pomeransky:2006bd,Elvang:2007rd,Iguchi:2007is,Evslin:2007fv,
Morisawa:2007di,Izumi:2007qx,Elvang:2007hs}.
It is quite important to study the stability of these black holes.  

The stability of higher dimensional Schwarzschild black holes
has been shown~\cite{Kodama:2003jz,Ishibashi:2003ap,Konoplya:2007jv}. 
For Myers-Perry black holes, the stability analysis is more difficult
 because of the difficulty of separation of gravitational perturbation equation.\footnote{The separability of geodesic Hamilton-Jacobi
equations, Klein-Gordon equations and Dirac equations in the Myers-Perry
spacetime have been shown 
in Refs.~\cite{Gibbons:1999uv,Frolov:2002xf,Frolov:2003en,Kunduri:2005fq,Kunduri:2005zg,Frolov:2006ib,Kubiznak:2006kt,Page:2006ka,Frolov:2006pe,Krtous:2006qy,Krtous:2007xf,Houri:2007uq,Houri:2007xz,Oota:2007vx,Sergyeyev:2007gf,Frolov:2007cb,Frolov:2008jr}.}
Nevertheless, there have been several works on the stability analysis. 
In the case of Myers-Perry black holes with equal angular momenta
 in odd dimensions higher than five, it was shown that 
 the special modes can be reduced to single ordinary differential equations. 
For these specific modes, the stability has been shown~\cite{Kunduri:2006qa}.
In the case of five-dimensional Myers-Perry black holes
with equal angular momenta, 
we have developed a method of analyzing the stability
by focusing on the spacetime symmetry $U(2)$~\cite{Murata:2007gv}. 
This  method has been proved to be useful for the stability analysis
of other $U(2)$ symmetric black holes~\cite{Kimura:2007cr,Ishihara:2008re}. 
The purpose of  this work is to  study the stability of five-dimensional Myers-Perry
black holes with equal angular momenta by making use of our method. 
We extend the previous stability analysis of Myers-Perry black holes
with equal angular momenta~\cite{Bizon:2007zf}.

The organization of this paper is as follows. 
In \S\ref{sec:MPBH}, we introduce Myers-Perry black holes
with equal angular momenta and discuss the spacetime symmetry. 
In \S\ref{sec:formalism}, we explain how to find master
variables with which one can deduce single ordinary equations. 
In \S\ref{sec:K=0stability}, the master equation for most symmetric
mode is derived. 
The stability for this case is analytically shown. 
In \S\ref{sec:MPK>0}, the master equations for other modes are
derived and the stability of Myers-Perry
black holes for these modes is numerically shown.
The final section is devoted to conclusions.
In Appendices \ref{app:eqsK=0} and \ref{app:drb_master_K=1}, 
detailed calculations for deriving master equations are shown. 


\section{Myers-Perry black holes with equal
 angular momenta and their symmetry}\label{sec:MPBH}

In this section, we introduce five-dimensional Myers-Perry  black holes with equal
angular momenta and clarify the symmetry of the spacetime. 

The metric is given by\footnote{
In our previous work~\cite{Murata:2007gv}, we used a
different coordinate. To put the metric into this form, we need the coordinate
transformation $r^2+a^2\rightarrow r^2$ and a change of the parameter $m\rightarrow 2\mu$.}
\begin{equation}
   ds^2= - dt^2
+\frac{d r^2}{G( r)}+
\frac{ r^2}{4} \{4\sigma^+\sigma^-+(\sigma^3)^2\}
+\frac{2\mu}{ r^2} \left(dt+\frac{a}{2}\sigma^3\right)^2\ ,
\label{eq:MPBH}
\end{equation}
where 
\begin{equation}
G( r) = 1 - \frac{2\mu}{ r^2}+\frac{2\mu a^2}{ r^4}\ ,
\end{equation}
and parameters $\mu$ and $a$ are defined for $\mu\geq 0$ and 
$a^2\leq \mu/2$.
Here, we define the invariant forms $\sigma^a\,(a=1,2,3)$ of $SU(2)$
satisfying the relation 
$d\sigma^a = 1/2 \epsilon^{abc} \sigma^b \wedge \sigma^c$ as
\begin{align}
  \sigma^1 &= -\sin\psi d\theta + \cos\psi\sin\theta d\phi\ ,\nonumber\\
  \sigma^2 &= \cos\psi d\theta + \sin\psi\sin\theta d\phi\ ,\nonumber\\
  \sigma^3 &= d\psi + \cos\theta d\phi  \ ,
\end{align}
and made a combination
\begin{equation}
 \sigma^{\pm} = \frac{1}{2}(\sigma^1 \mp i \sigma^2)\  .
\end{equation}
The coordinate ranges are
 $0\leq \theta < \pi $, $0\leq \phi <2\pi$, and $0\leq \psi <4\pi$. 
The dual vectors of $\sigma^a$ are given by
\begin{align}
 {e}_1 &= -\sin\psi \partial_\theta +
  \frac{\cos\psi}{\sin\theta}\partial_\phi - \cot\theta\cos\psi
 \partial_\psi\ ,\nonumber\\
 {e}_2 &= \cos\psi \partial_\theta +
  \frac{\sin\psi}{\sin\theta}\partial_\phi - \cot\theta\sin\psi
 \partial_\psi\ ,\nonumber\\
 {e}_3 &= \partial_\psi\ ,
\end{align}
and, by definition, they satisfy the relation $\sigma^a_i {e}^i_b = \delta^a_b$. 

The horizon $r=r_+$ is found by solving $G(r_+)=0$, namely, 
\begin{equation}
 r_+^2=\mu+\sqrt{\mu(\mu-2a^2)}\ .
\label{eq:r_+}
\end{equation}
The angular velocity of the horizon is given by 
\begin{equation}
 \Omega_H=\frac{a}{r_+^2}\ .
\label{eq:degen_angular_vel}
\end{equation}
In term of $ r_+$ and $\Omega_H$, the two parameters $(a,\mu)$ in the
metric~(\ref{eq:MPBH}) can be rewritten as
\begin{equation}
a = r_+^2 \Omega_H\ ,\quad
\mu=
 \frac{1}{2}\frac{ r_+^2}{1-\Omega_H^2r_+^2}\ .  
\label{eq:mu_a_to_r+_Omega}
\end{equation}
We see the the upper bound of the angular velocity,
\begin{equation}
 \Omega_H \leq
  \frac{1}{\sqrt{2}r_+}
 \equiv \Omega_H^\text{max}\ .
\label{eq:Omegamax}
\end{equation}

Apparently, the metric~(\ref{eq:MPBH}) has the $SU(2)$ symmetry
characterized by Killing vectors $\xi_\alpha \ , (\alpha=x,y,z)$:
\begin{align}
 \xi_x &= \cos\phi\partial_\theta +
 \frac{\sin\phi}{\sin\theta}\partial_\psi -
 \cot\theta\sin\phi\partial_\phi\ ,\nonumber\\
 \xi_y &= -\sin\phi\partial_\theta +
 \frac{\cos\phi}{\sin\theta}\partial_\psi -
 \cot\theta\cos\phi\partial_\phi\ ,\nonumber\\
 \xi_z &= \partial_\phi\ .
\end{align}
The symmetry can be explicitly shown by using
 the relation $\mathcal{L}_{\xi_\alpha}\sigma^a=0$,
 where $\mathcal{L}_{\xi_\alpha}$ is a Lie derivative
 along the curve generated by the vector field $\xi_\alpha$.
From the metric (\ref{eq:MPBH}), we can also read off 
the additional $U(1)$ symmetry, which 
keeps part of the metric, $\sigma^+\sigma^-$, invariant. 
Thus, the symmetry of
five-dimensional degenerate Myers-Perry black hole 
becomes $SU(2)\times U(1) \simeq U(2)$.\footnote{
The spacetime~(\ref{eq:MPBH}) also has a time translation
symmetry generated by $\partial/\partial t$. Because of this symmetry, we can
separate the  time dependence of fields as $\propto e^{-i\omega
t}$. However, this is obvious and we will not pay much attention to this
symmetry hereafter.}
The Killing vectors $  {e}_3 , \xi_x , \xi_y$ and $\xi_z$ constitute the symmetry, 
i.e., $ {e}_3$ is a generator of $U(1)$ and $\xi_\alpha\,(\alpha=x,y,z)$
are generators of $SU(2)$.

Let us define the two kinds of angular momentum operators
\begin{equation}
  L_\alpha = i \xi_\alpha \ , \quad
  W_a  = i  {e}_a \ ,
\end{equation}
where $\alpha,\beta,\cdots = x,y,z$ and $a,b,\cdots = 1,2,3$. 
They satisfy the commutation relations
\begin{equation}
 [L_\alpha, L_\beta] = i \epsilon_{\alpha\beta\gamma} L_\gamma \  , \quad
 [W_a, W_b] = -i\epsilon_{abc} W_c
 \ , \quad [L_\alpha, W_a]=0 \ ,
\end{equation}
where $\epsilon_{\alpha\beta\gamma}$ and $\epsilon_{abc}$ are
antisymmetric tensors that satisfy $\epsilon_{123}=\epsilon_{xyz}=1$. 
Note that $L^2 \equiv L_\alpha^2 = W_a^2$.
The symmetry group, $U(2)\simeq SU(2)\times U(1)$, is generated by $L_\alpha$ and
$W_3$.
Here, we should note the fact that
\begin{eqnarray}
\mathcal{L}_{W_3} \sigma^{\pm}
=  \pm  \sigma^{\pm} \ ,\quad 
\mathcal{L}_{W_3} \sigma^3= 0\ .
 \label{rule}
\end{eqnarray}

Let us construct the representation of  $U(2)$.
The eigenfunctions of $L^2$ are degenerate, but
can be completely specified by eigenvalues of other operators $L_z$ and $W_3$.
They are called Wigner functions and are defined as 
\begin{equation}
 L^2 D^J_{KM} = J(J+1)D^J_{KM}\ ,\quad
 L_z D^J_{KM} = M D^J_{KM}\ ,\quad
 W_3 D^J_{KM} = K D^J_{KM}\ ,
\label{eq:WigDef}
\end{equation}
where $J=0,1/2,1,3/2\cdots$ and $M,K=-J,-J+1,\cdots,J$. 
From Eqs.~(\ref{eq:WigDef}), 
we see that $D^J_{KM}$ forms the irreducible
representation of $U(2)\simeq SU(2) \times U(1)$.
The Wigner functions are functions of $(\theta,\phi,\psi)$ and
 satisfy the orthonormal relation:
\begin{equation}
 \int^\pi_0 d\theta \int^{2\pi}_0 d\varphi \int^{4\pi}_0 d\psi
 \sin\theta\, D^J_{KM}(\theta,\phi,\psi)D^{J'\,\ast}_{K'M'}{}(\theta,\phi,\psi) =
 \delta_{JJ'}\delta_{KK'}\delta_{MM'}\ .
\end{equation}
The following relations are useful for later calculations
\begin{equation}
 W_+ D^J_{KM} = i \epsilon_K D^J_{K-1,M}\ ,\quad W_- D^J_{KM} = -i\epsilon_{K+1}D^J_{K+1,M}\
  , \quad W_3 D^J_{KM} = KD^J_{KM}\ ,
\end{equation}
where we have defined $W_\pm = W_1\pm i W_2$ and $\epsilon_K=\sqrt{(J+K)(J-K+1)}$. 
From this relation, we obtain the differential rule of Wigner
functions as
\begin{equation}
 \partial_+ D^J_{KM} = \epsilon_K D^J_{K-1,M}\ ,\quad
 \partial_- D^J_{KM} = -\epsilon_{K+1}D^J_{K+1,M}\ ,\quad
 \partial_3 D^J_{KM} = -iKD^J_{KM}\ ,
\label{eq:delD}
\end{equation}
where we have defined 
$\partial_\pm \equiv e_\pm^i\partial_i$ and $ \partial_3 \equiv e_3^i\partial_i$.

\section{A way to find master variables}\label{sec:formalism}

For the stability analysis of the Myers-Perry black hole~(\ref{eq:MPBH}),
it is necessary to find master equations for metric perturbations. 
Because Myers-Perry  spacetime~(\ref{eq:MPBH}) has the 
$U(2)$ symmetry, a group theoretical method with a twist can be used 
~\cite{Murata:2007gv,Hu:1974hh}. 
In this section, we explain a way to find master variables. 
Once the master variables are found,
 we can deduce the master equations from the gravitational perturbation equations
\begin{multline}
 \delta G_{\mu\nu}=\frac{1}{2}\big[\nabla^\rho\nabla_\mu h_{\nu \rho}
 + \nabla^\rho\nabla_\nu h_{\mu\rho} -
  \nabla^2 h_{\mu\nu} - \nabla_\mu\nabla_\nu h\\
-g_{\mu\nu}(\nabla^\rho\nabla^\sigma h_{\rho\sigma} - \nabla^2 h -
 R^{\rho\sigma}h_{\rho\sigma}) - R h_{\mu\nu}\big] = 0\ ,
\label{eq:tensor_EOM}
\end{multline}
where $\nabla_\mu$ denotes the covariant derivative with respect to $g_{\mu\nu}$,
and we have defined $g_{\mu\nu}\rightarrow g_{\mu\nu}+h_{\mu\nu}$
and $h= g^{\mu\nu} h_{\mu\nu}$.

Now, we consider the mode expansion of $h_{\mu\nu}$. 
The metric perturbation can be classified into three parts,
$h_{AB},h_{Ai}$ and $h_{ij}$ where $A,B=t,r$ and $i,j=\theta,\phi,\psi$. 
They behave as a scalar, vector and tensor for
coordinate transformations of $\theta,\phi$ and $\psi$, respectively. 
The scalar
$h_{AB}$ can be expanded by Wigner functions immediately as 
\begin{equation}
 h_{AB} = \sum_{K} h_{AB}^K(x^A) D_K(x^i)\ .
\label{eq:hAB_dec}
\end{equation}
Here, we have omitted the indices $J$ and $M$ because the differential rule of Wigner
function~(\ref{eq:delD}) cannot shift $J$ and $M$ and therefore the modes with different
eigenvalues $J$ and $M$ are trivially decoupled in the perturbation equations. 

To expand the vector part $h_{Ai}$, we need an elaborate method. 
First, we change the basis $\{\partial_i\}$ to
$\{ {e}^a\}$, that is, $h_{Ai}=h_{Aa}\sigma^a_i$ where $a=\pm,3$. 
Then, because $h_{Aa}$ is scalar, we can expand it, using Wigner functions as
\begin{align}
 h_{Ai}(x^\mu)&=h_{A+}(x^\mu)\sigma^+_i + h_{A-}(x^\mu)\sigma^-_i +
 h_{A3}(x^\mu)\sigma^3_i \nonumber\\
 & = \sum_K \left[h_{A+}^K(x^A) \sigma^+_i D_{K-1} + h_{A-}^K(x^A)\sigma^-_i D_{K+1}
 + h_{A3}^K(x^A)\sigma^3_iD_K\right]\ .
\label{eq:hAi_dec}
\end{align}
In the expansion of $h_{A+}$, $h_{A-}$ and $h_{A3}$, we
shift the index $K$ of Wigner functions, for example, 
$h_{A+}$ is expanded as $\sum_K h_{A+}^K D_{K-1}$. 
The reason is as follows. 
The invariant forms $\sigma^\pm$ and 
$\sigma^3$ have the $U(1)$ charge $\pm1$ and $0$, respectively 
(see Eq.~(\ref{rule})), while
Wigner function $D_K$ has the $U(1)$ charge $K$ (see
Eq.~(\ref{eq:WigDef})). 
Therefore, by shifting the index $K$, we can assign the same $U(1)$ charge $K$
to $\sigma^+_i D_{K-1}$, $\sigma^-_i D_{K+1}$ and $\sigma^3_iD_K$ in
Eq.~(\ref{eq:hAi_dec}). 

The expansion of tensor part $h_{ij}$ can be carried out in a
similar way  as
\begin{eqnarray}
 h_{ij}(x^\mu) 
 &=&  \sum_K  \left[ h_{++}^K  \sigma^+_i \sigma^+_j  D_{K-2} 
  +   2 h_{+-}^K\sigma^+_i \sigma^-_j D_{K} 
  +  2 h_{+3}^K\sigma^+_i \sigma^3_j D_{K-1}   \right. \nonumber \\
&& \left.  \qquad   +  h_{--}^K \sigma^-_i \sigma^-_j D_{K+2} 
  + 2 h_{-3}^K\sigma^-_i \sigma^3_j D_{K+1} 
  +   h_{33}^K \sigma^3_i \sigma^3_j D_{K}    \right]     \  .
\label{eq:hij_dec}
\end{eqnarray}
To assign the same $U(1)$ charge $K$  to each term, 
we shift the index $K$ of Wigner functions. 

Substituting Eqs.~(\ref{eq:hAB_dec}), (\ref{eq:hAi_dec}) and 
(\ref{eq:hij_dec}) into the perturbation equations~(\ref{eq:tensor_EOM}),
 we obtain the equations for each mode labelled by $J$, $M$ and $K$. 
Because of $U(2)$ symmetry, different eigenmodes cannot appear in the
same equation.  

It is interesting that we can find master variables from the above information. 
First, we should note that coefficients of the expansion have different indices $K$ 
and, therefore, coefficients of components $h_{AB}^K$, $h_{Aa}^K$ and $h_{ab}^K$ are 
restricted as follows:
\begin{equation}
\begin{array}{|c|c|c|c|c|}
\hline
h_{++}      & h_{A+},h_{+3} & h_{AB}, h_{A3} , h_{+-},h_{33}&h_{A-},h_{-3} &h_{--} \\ \hline
|K-2|\leq J &|K-1|\leq J    &|K|\leq J                      &|K+1|\leq J   &|K+2|\leq J \\ \hline
\end{array}
\nonumber
\end{equation}
For example,  for the $J=0$ mode,  we can classify the metric perturbation as follows:
\begin{equation}
\begin{array}{|c|c|c|c|c|}
\hline
h_{++} & h_{A+},h_{+3} & h_{AB}, h_{A3} , h_{+-},h_{33}&h_{A-},h_{-3} &h_{--} \\ \hline
K=2    &               &                        &              &       \\ \hline
       &K=1            &                        &              &       \\ \hline
       &               &K=0                     &              &       \\ \hline
       &               &                        &K=-1          &       \\ \hline
       &               &                        &              &K=-2   \\ \hline
\end{array}
\nonumber
\end{equation}
In the above table, variables in each row can couple with each other.
Apparently, $h_{\pm\pm}$ are decoupled, hence, it is straightforward
to obtain the master equation for these variables.
Other variables,
$(h_{A+},h_{+3})$, $(h_{AB}, h_{A3} , h_{+-},h_{33})$ and $(h_{A-},h_{-3})$,
are coupled in each set. However, after fixing the gauge degrees of freedom,
we have the master equation for each set. In total, there are five master equations,
which matches the physical degrees of freedom of the tensor perturbations at this level.
We can continue this analysis. 
For $J=1/2$ modes, we have the following table:
\begin{equation}
\begin{array}{|c|c|c|c|c|}
\hline
h_{++} & h_{A+},h_{+3} & h_{AB}, h_{A3} , h_{+-},h_{33}&h_{A-},h_{-3} &h_{--} \\ \hline
K=5/2  &               &                        &              &       \\ \hline
K=3/2  &K=3/2          &                        &              &       \\ \hline
       &K=1/2          &K=1/2                   &              &       \\ \hline
       &               &K=-1/2                  &K=-1/2        &       \\ \hline
       &               &                        &K=-3/2        &K=-3/2 \\ \hline
       &               &                        &              &K=-5/2 \\ \hline
\end{array}
\nonumber
\end{equation}
From the above analysis, we  see that 
$(J,M,K=\pm(J+2))$ modes are always  decoupled.
Each mode can be reduced to the single master equation. 
Thus, we have shown that we  obtain an infinite number of master equations
for the metric perturbations, although they are not everything. 

Because of the relations $h_{--}^\ast=h_{++}$, 
$h_{A-}^\ast=h_{A+}$ and $h_{-3}^\ast=h_{+3}$, 
we see that $(J,M,-K)$ modes are complex conjugate of $(J,M,K)$
modes. Therefore, we will assume $K\geq 0$ in the following sections.

\section{Stability analysis for $(J=0,M=0,K=0)$ mode}\label{sec:K=0stability}
In the previous section, we showed how to find master variables. 
In this section, we will derive the master equation for the $J=M=K=0$ mode
and show the stability for this mode.
The stability for this mode has previously been shown in Ref.~\cite{Bizon:2007zf}. 
However, we will show the stability again using our formalism. 

For the $(J=0,M=0,K=0)$ mode, we must consider metric components
$h_{tt},h_{t r},h_{ r r},h_{t3},h_{ r3},h_{+-}$ and $h_{33}$. 
We set $h_{\mu\nu}$ as
\begin{multline}
 h_{\mu\nu}dx^\mu dx^\nu = \text{Re}\big\{
e^{-i\omega t}\big[h_{tt}(r) dt^2  + 2h_{t r}(r) dtd r + h_{ r r}(r) d r^2
+ 2h_{t3}(r) dt\sigma^3\\
 + 2h_{ r3}(r) d r\sigma^3 + 2h_{+-}(r) \sigma^+ \sigma^-
+ h_{33}(r) \sigma^3 \sigma^3\big]\big\}\ ,
\label{eq:h_K=0}
\end{multline}
where ${\rm Re}$ represents the real part of a complex quantity.
With the gauge parameters 
\begin{equation}
\begin{split}
 \xi_A(x^\mu) =  \text{Re}\{\xi_A(r) e^{-i\omega t}\}\ ,\quad
 \xi_i(x^\mu) =  \text{Re}\{\xi_3(r) e^{-i\omega t}\sigma^3_i\} \ ,
\end{split} 
\end{equation} 
the gauge transformations for those components are given by
\begin{align}
&\delta h_{tt}=
-2i\omega\xi_t
-\frac {4\mu G(r)}{r^3}\xi_r\ ,\nonumber\\
&\delta h_{tr}=
\xi_t'
-\frac{4\mu}{r^3G(r)}\xi_t
-i\omega\xi_r
+\frac{8\mu a}{r^5G(r)}\xi_3\ ,
\nonumber\\
&\delta h_{t3}=
-\frac{2\mu a G(r)}{r^3}\xi_r
-i\omega\xi_3\ ,\nonumber\\
&\delta h_{rr}=
2\xi_r'
+\frac{4\mu(r^2-2a^2)}{r^5 G(r)}\xi_r\ ,
\nonumber\\
&\delta h_{r3}=
-\frac{4\mu a}{r^3G(r)}\xi_t
+\xi_3'
-\frac{2(r^4-2\mu r^2-2\mu a^2)}{r^5G(r)}\xi_3\ ,
\nonumber\\
&\delta h_{+-}=rG(r)\xi_r\ ,\nonumber\\
&\delta h_{33}=
\frac{(r^4-2\mu a^2)G(r)}{2r^3}\xi_r\ ,
\end{align}
where $'\equiv \partial_r$ and 
$\delta$ represents the gauge transformation 
defined by 
$\delta h_{\mu\nu}=\nabla_\mu \xi_\nu+ \nabla_\nu\xi_\mu$.
The following gauge conditions fix the gauge degrees of freedom completely:
\begin{equation}
h_{tt}=0\ ,\quad
h_{t3}=0\ ,\quad h_{33}=0\ .
\label{eq:K=0_gauge_choice}
\end{equation}

Substituting Eqs.~(\ref{eq:h_K=0}) and (\ref{eq:K=0_gauge_choice})
into Eq.~(\ref{eq:tensor_EOM}), we obtain a set of ordinary
differential equations. These equations can be found  in
Appendix~\ref{app:eqsK=0}. Eliminating  $h_{tr},h_{rr}$ and $h_{r3}$
from these equations, we obtain the Schr\"{o}dinger-type master equation, 
\begin{equation}
 -\frac{d^2\Phi_0}{d r_\ast^2} + V_0( r)\Phi_0 = \omega^2 \Phi_0\ ,
\label{eq:K=0_master}
\end{equation}
where we define the new variable
\begin{equation}
 \Phi_0\equiv 
\frac {( r^4-2\mu a^2)({ r}^{4}+2\mu a^2)^{1/4}}
{ r^{3/2}(3 r^4+2\mu a^2)}h_{+-}\ ,
\end{equation}
and the tortoise coordinate 
\begin{equation}
 d r_\ast = \frac{( r^4+2\mu a^2)^{1/2}}{G( r) r^2}dr\ .
\label{eq:kame}
\end{equation}
The potential $V_0$ is given by
\begin{multline}
V_0( r) =
 \frac{G(r)}{4(3 r^4+2\mu a^2)^2( r^4+2\mu a^2)^3 r^2}
 \big[
315 r^{20}
+162\mu r^{18}
+2430\mu a^2  r^{16}\\
+1392\mu^2 a^2 r^{14}
+5400\mu^2 a^4  r^{12}
+5808 \mu^3 a^4  r^{10}
+2608\mu^3 a^6  r^8\\
+6080\mu^4 a^6  r^6
-2064\mu^4 a^8  r^4
+32\mu^5 a^8  r^2
-160\mu^5 a^{10}
\big]\ .
\label{eq:K=0pot}
\end{multline}

\begin{figure}
\begin{center}
\includegraphics[scale=0.5]{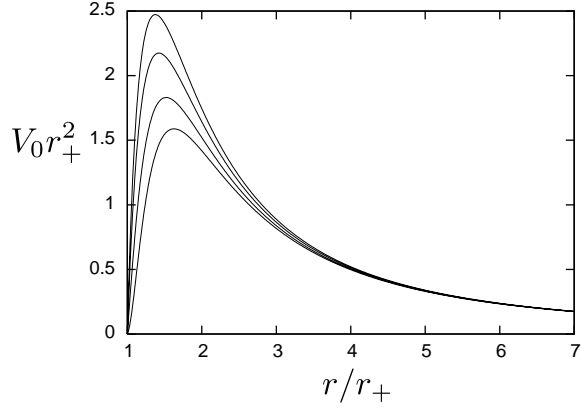}
\caption{\label{fig:potential_V0_AdS}Typical profiles for the potential
 $V_0$ are depicted.  From top to
 bottom, each curve represents the potential for 
 $\Omega_H/\Omega_H^\text{max}=0.1,0.7,0.9$ and $0.99$. We see the positivity
 of these potentials.}
\end{center}
\label{fig:pot}
\end{figure}

To prove the stability, we must show the positivity of $V_0$.
The typical profiles of $V_0$ shown in
Fig.~\ref{fig:potential_V0_AdS} indicate that the potential is  always positive.
 In fact, the positivity can be proved from the 
expression~(\ref{eq:K=0pot}). 
To see the positivity of $V_0(r)$, we focus on $r^6$, $r^4$ and $r^0$ terms
in the big bracket in Eq.~(\ref{eq:K=0pot}).
After dividing them by $16 \mu^4 a^6$, we collect them as
\begin{equation}
f(r)=380r^6-129a^2r^4-10\mu a^4\ .
\label{eq:g}
\end{equation}
If $f(r)$ is positive, $V_0(r)$ is also positive. 
From Eq.~(\ref{eq:r_+}),  we see
$\mu\leq r_+^2$ and $a^2\leq r_+^2/2$. 
Therefore, 
\begin{equation}
 f(r) \geq 380r^6-\frac{129}{2} r_+^2 r^4-\frac{5}{2}r_+^6
=313r^6 + \frac{129}{2}r^4(r^2-r_+^2) + \frac{5}{2}(r^6-r_+^6)>0\ .
\end{equation}
This proves  the stability for the $J=M=K=0$ mode. 

\section{Stability analysis for $K\neq 0$ modes}\label{sec:MPK>0}

\subsection{Master equation for $(J=0,M=0,K=1)$ mode}

As we have shown in \S\ref{sec:formalism}, for the $(J=0,M=0,K=1)$ mode,
we have to consider components $h_{t+},h_{ r+}$ and $h_{+3}$, namely,  
\begin{multline}
 h_{\mu\nu}dx^\mu dx^\nu = \text{Re}\big\{
2h_{t+}(t, r) dt\,\sigma^+ + 2h_{ r+}(t, r) d r\,\sigma^+ +
2 h_{+3}(t, r) \sigma^+ \sigma^3 \big\}\ .
\label{eq:h_K=1}
\end{multline}
With the gauge parameter
$\xi_i(x^\mu)=\text{Re}\{\xi_+(t,r)\sigma^+_i\}$, 
the gauge transformations for these components are given by
\begin{equation}
\delta h_{t+} =
 \dot{\xi}_+ + \frac{4i\mu a}{ r^4}\xi_+\ ,\quad
\delta h_{ r+} =
\xi_+'-\frac{2}{ r}\xi_+\ ,\quad
\delta h_{+3} =
\frac{2i\mu a^2}{ r^4}\xi_+\ ,
\end{equation}
where $^\cdot\equiv \partial/\partial t$ and 
$' \equiv \partial/\partial r$. 
To derive the master equation, it is convenient to use the action
  instead of equations of motion~(\ref{eq:tensor_EOM}). 
The action for metric perturbation in a vacuum is given by
\begin{multline}
 S = \frac{1}{4}\int d^5x \sqrt{-g}\,[-\nabla_\mu
 h_{\nu\rho} \nabla^\mu h^{\nu\rho} + \nabla_\mu h \nabla^\mu h \\
+2\nabla_\mu h_{\nu\rho} \nabla^\nu h^{\rho\mu} - 2\nabla^\mu h_{\mu\nu}
 \nabla^\nu h] \ ,
\label{eq:grav_action}
\end{multline}
where we use the unit $16\pi G_5=1$; here $G_5$ is the five-dimensional
Newton's constant. 
Now, we define gauge invariant variables,
\begin{equation}
 f_{t}\equiv
\frac{h_{t+}}{ r^2} + \frac{i r^2}{2\mu a^2}\dot{h}_{+3} - \frac{2}{a r^2}h_{+3}
\ ,\quad
 f_{ r}\equiv 
\frac{h_{ r+}}{ r^2} + \frac{i}{2\mu a^2}( r^2 h_{+3})'
\ .
\end{equation}
In term of these variables, the action~(\ref{eq:grav_action}) becomes
\begin{multline}
 S=\frac{1}{4}\int dtd r \bigg[
\frac{ r( r^4+2\mu a^2)}{4}|\dot{f}_ r-f_t'|^2 
+ \frac{4\mu^2 a^4}{ r^5 G(r)}|f_t|^2 \\
+ \frac{4\mu^2a^2(r^2 -a^2)}{ r^5}|f_ r|^2
+2\mu a\,\text{Im}\big\{ r(\dot{f}_ r-f_t')f_ r^\ast - 4f_t f_ r^\ast\big\}
\bigg]\ .
\label{eq:action_K=1}
\end{multline}
In the above action, there are two fields $f_t$ and $f_ r$. However, $f_t$
is not the physical degree of freedom. 
Therefore, we can eliminate it from the action and get the master equation. 
In Appendix~\ref{app:drb_master_K=1}, we show the details. 
As a result, in terms of the new variable
\begin{equation}
 \Phi_1 = 
\frac{ r^{3/2}( r^4+2\mu a^2)^{1/4}}
{( r^{10}+2\mu a^2 r^6+\mu^2 a^6)^{1/2}}
\bigg[
\frac{ r( r^4+2\mu a^2)}{4}(\dot{f}_ r-f_t')+i\mu a r f_ r
\bigg]\ ,
\end{equation}
the master equation can be obtained as
\begin{equation}
 -\frac{d^2 \Phi_1}{d r_\ast^2} + V_1( r)\Phi_1 =
  [\omega-2\Omega_1( r)]^2 \Phi_1\ ,
\label{eq:master_K=1}
\end{equation}
where
\begin{equation}
 \Omega_1( r) = \frac{2\mu a}{ r^4+2\mu a^2}
\left(1-\frac{a^2  r^4 (5 r^4+6\mu a^2)G(r)}{4( r^{10}+2\mu a^2  r^6+ \mu^2a^6)} \right)
\ ,
\end{equation}
and
\begin{align}
V_1( r)=&
\frac{G(r)}{4 r^2(r^4+2\mu a^2)^3( r^{10}+2\mu a^2 r^6+\mu^2a^6)^2}
\big[
35r^{32}
+18\mu r^{30}
+310\mu a^2 r^{28}\nonumber\\
&+160\mu^2 a^2 r^{26}
+1192\mu^2 a^4 r^{24}
+2\mu^2 a^4(152\mu-75a^2)r^{22}
+3068\mu^3 a^6 r^{20}\nonumber\\
&-64\mu^3 a^6(2\mu+15a^2)r^{18}
+5208\mu^4 a^8 r^{16}
-16\mu^4 a^8(30\mu+133a^2)r^{14}\nonumber\\
&+3\mu^4 a^{10}(1424\mu+5a^2)r^{12}
-1654\mu^5 a^{12} r^{10}
+2\mu^5 a^{12}(432\mu+25a^2)r^8\nonumber\\
&-168\mu^6 a^{14} r^6
+68\mu^6 a^{16} r^4
-24\mu^7 a^{16} r^2
+56\mu^7 a^{18}\big]\ .
\end{align}
We used the tortoise coordinate defined in
Eq.~(\ref{eq:kame}). 
In Eq.~(\ref{eq:master_K=1}), 
we separated the $t$ dependence of $\Phi_{1}$ by the  Fourier transformation 
$\Phi_{1}(t, r) = e^{-i\omega t}\Phi_{1}( r)$. 

The asymptotic forms of $\Omega_1(r)$ and $V_1( r)$ are
\begin{equation}
\Omega_1( r) \rightarrow 0 \quad( r\rightarrow \infty)\ ,\qquad
\Omega_1( r) \rightarrow \Omega_H\ , \quad( r\rightarrow  r_+)  
\end{equation}
and
\begin{equation}
 V_1( r)\rightarrow 0\ , \quad( r\rightarrow  r_+, \infty)
\end{equation}
where $\Omega_H$ is the angular velocity of the horizon defined
in Eq.~(\ref{eq:degen_angular_vel}). 
Therefore, the asymptotic form of the 
solution of master equation~(\ref{eq:master_K=1}) becomes
\begin{equation}
\Phi_1 \rightarrow e^{\pm i\omega  r_\ast}\ ,
 \quad( r\rightarrow \infty)\ ,\quad
 \Phi_1 \rightarrow e^{\pm i(\omega-2\Omega_H) r_\ast}\ .
 \quad( r\rightarrow  r_+)
\end{equation}
Before discussing the stability for this mode,
we  shall derive the master equation for $(J,M,K=J+2)$ modes. 
The stability analyses of both cases will be discussed simultaneously.

\subsection{Master equations for $(J,M,K=J+2)$ modes}
To obtain the master equations for $(J,M,K=J+2)$ modes, we set $h_{\mu\nu}$ as
\begin{equation}
 h_{\mu\nu}(x^\mu) dx^\mu dx^\nu 
 = {\rm Re} \left[ h_{++}(t,r)D_{J}(x^i)\sigma^+ \sigma^+ \right]  \ ,
\label{eq:h_K=J+2}
\end{equation}
where $D_J\equiv D_{K=J,M}^J$. 
This $h_{++}$ field is gauge invariant. 
Substituting Eq.~(\ref{eq:h_K=J+2}) into Eq.~(\ref{eq:grav_action}) and
using the differential rule of Wigner functions~(\ref{eq:delD}), we 
obtain the action for the $K=J+2$ mode,
\begin{multline}
S =  \frac{1}{4}  \int dtdr \bigg[
\frac{ r^4+2\mu a^2}{4 r^5 G(r)}|\dot{h}_{++}|^2 -
 \frac{G(r)}{4 r}|h_{++}'|^2
-\frac{i(J+2)\mu a}{ r^5 G(r)}(\dot{h}_{++}h_{++}^\ast-\dot{h}_{++}^\ast
 h_{++})\\
+\frac{1}{ r^{11} G(r)}
\big\{
-(J+1)(J+2) r^8
+2\mu(J^2+3J+1) r^6\\
+2\mu(2\mu +(J+4)a^2) r^4
-12\mu^2 a^2 r^2
+8\mu^2 a^4
\big\}
|h_{++}|^2
\bigg]\ ,
\end{multline}
where the asterisk denotes the complex conjugate. 
We can derive the equations of motion for $(J,M,K=J+2)$ modes from 
the above action. 
Defining the new variable
\begin{equation}
 \Phi_{J+2} = \frac{( r^4+2\mu a^2)^{1/4}}{ r^{3/2}}h_{++}\ ,
\end{equation}
we obtain the master equation,
\begin{equation}
 -\frac{d^2 \Phi_{J+2}}{d r_\ast^2} + V_{J+2}( r)\Phi_{J+2} =
  [\omega-2(J+2)\Omega_{J+2}( r)]^2 \Phi_{J+2}\ ,
\label{eq:master_K=J+2}
\end{equation}
where 
\begin{equation}
 \Omega_{J+2}( r)=\frac{2\mu a}{ r^4 + 2\mu a^2} ,
\end{equation}
and 
\begin{multline}
V_{J+2}( r) = \frac{G( r)}{4 r^2( r^4+2\mu a^2)^3}
[
(4J+7)(4J+5) r^{12}
+ 18\mu r^{10}
+ 2\mu a^2 (16J^2+32J+5) r^8\\
-40\mu^2 a^2 r^6
-4\mu^2 a^4(16J+35) r^4
+8\mu^3 a^4 r^2
-40\mu^3 a^6
]\ .
\end{multline}
We separated the $t$ dependence of $\Phi_{J+2}$ by the  Fourier transformation 
$\Phi_{J+2}(t, r) = e^{-i\omega t}\Phi_{J+2}( r)$. 

Since the asymptotic forms of $\Omega_{J+2}(r)$ and $V_{J+2}( r)$
become 
\begin{equation}
\Omega_{J+2}( r) \rightarrow 0 \quad( r\rightarrow \infty)\ ,\qquad
\Omega_{J+2}( r) \rightarrow \Omega_H \ ,\quad( r\rightarrow  r_+)  
\end{equation}
and
\begin{equation}
 V_{J+2}( r)\rightarrow 0 \ ,\quad( r\rightarrow  r_+, \infty)
\end{equation}
we get the asymptotic form of
solution of master equation~(\ref{eq:master_K=J+2})  as
\begin{equation}
 \Phi_{J+2} \rightarrow e^{\pm i\omega  r_\ast}\ ,
 \quad( r\rightarrow \infty)\ ,\quad
 \Phi_{J+2} \rightarrow e^{\pm i\{\omega-2(J+2)\Omega_H\} r_\ast}\ .
 \quad( r\rightarrow  r_+)
\end{equation}

\subsection{Method to study the stability of $K\neq 0$ modes}\label{sec:MP_K>0}

Since the master equations for $K\neq 0$ modes are not of the Schr\"{o}dinger type, 
we cannot show the stability for
these modes from the positivity of the potential  as was done in 
the case of the $J=M=K=0$ mode. 
Here, we will follow the method used to show the stability of Kerr black
holes~\cite{Press:1973,Kunduri:2006qa}. 

To discuss the master equations for $(J=0,M=0,K=1)$ and $(J,M,K=J+2)$
simultaneously,  we  write the master equations as
\begin{equation}
-\frac{d^2 \Phi_K}{d r_\ast^2} + V_K( r)\Phi_K =
  [\omega-2K\Omega_K( r)]^2 \Phi_K\ .
\label{eq:master_unify}
\end{equation}
For $K=1$ and $K=J+2$, the 
above master equations reduce to Eq.~(\ref{eq:master_K=1}) and
Eq.~(\ref{eq:master_K=J+2}), respectively. 

Now, we will examine if there exists a quasi-normal mode with 
$\text{Im}\,\omega>0$ in the system. 
Because of the time dependence $h_{\mu\nu} \propto e^{-i \omega t}$,
the existence of such a mode implies instability of the system. 
Recall that the boundary conditions for quasi-normal modes  are given by
\begin{align}
&\Phi_K \rightarrow e^{- i(\omega-2K\Omega_H) r_\ast}\ ,
 \quad( r\rightarrow  r_+)\nonumber\\
&\Phi_K \rightarrow 
Z_\text{out}e^{i\omega  r_\ast}\ .
\quad( r\rightarrow \infty)
\label{eq:MPbc}
\end{align}
Namely, the wave function must be ingoing at the
horizon and outgoing at infinity.

To study the stability using the master equation~(\ref{eq:master_unify}), 
we start with the assumption that  Myers-Perry black holes are stable
for sufficiently small angular velocity $\Omega_H$. This is
a natural assumption because the higher dimensional Schwarzschild black hole
is stable~\cite{Ishibashi:2003ap}. In fact, in the limit of
$a\rightarrow 0$, the master equation~(\ref{eq:master_unify}) takes the 
Schr\"{o}dinger form and the positivity of the potential is easily
seen. 
Under this assumption, for small $\Omega_H$, 
the imaginary part of the quasi-normal frequency must be negative, 
$\text{Im}\,\omega < 0$. 
Now, if there exists instability, a quasi-normal mode with 
$\text{Im}\,\omega>0$ will appear at some point as we increase $\Omega_H$. 
This means that
one of quasi-normal modes must cross the real
axis in the complex $\omega$ plane for some $\Omega_H$. 
Therefore, if the black hole is unstable for large $\Omega_H$, there must be 
a critical value $\Omega_H=\Omega_H^\text{crit}$ for which there exists a mode
with $\text{Im}\,\omega=0$ under the  boundary condition (\ref{eq:MPbc}). 
We look for such $\Omega_H^\text{crit}$. 

For the numerical analysis,
 it is convenient to define quasi-normal modes as follows.
 Given the ingoing wave at the horizon,
\begin{equation}
\Phi_K \rightarrow e^{- i(\omega-2K\Omega_H) r_\ast}\ ,
 \quad( r\rightarrow  r_+)
\label{eq:MPbc1}
\end{equation}
 we generally obtain the wave function at infinity as 
\begin{equation}
 \Phi_{K} \rightarrow Z_\text{out} e^{i\omega  r_\ast} + Z_\text{in}
 e^{-i\omega  r_\ast}\ ,
 \quad( r\rightarrow \infty)
\label{eq:MPbc2}
\end{equation}
where $Z_\text{out}$ and $Z_\text{in}$ are constants. 
Hence,  quasi-normal modes are defined by the condition $Z_\text{in}=0$.

For the purpose of searching for $\Omega_H^\text{crit}$, we take $\omega$
to be real. In this case, the Wronskian of $\Phi_{K}$ is conserved, that is,
\begin{equation}
 \text{Im}\left[\Phi^\ast_{K}\frac{d}{d r_\ast}\Phi_{K}\right]^{ r= r_2}_{ r= r_1}
 = 0\ ,
\label{eq:Wronskian}
\end{equation}
for any $ r_1$ and $ r_2$. We take $ r_1= r_+$ and $ r_2=\infty$.  
Then, from Eq.~(\ref{eq:Wronskian}), we obtain the relation,
\begin{equation}
 2K\Omega_H - \omega = \omega|Z_\text{out}|^2\ ,
\label{eq:in_out_rel}
\end{equation}
where we used the boundary condition~(\ref{eq:MPbc}).
Since the master equation~(\ref{eq:master_unify}) is invariant under 
$\omega\rightarrow -\omega$ and $a\rightarrow -a$,  we take
$\omega\geq 0$. 
Then, from Eq.~(\ref{eq:in_out_rel}), we obtain the inequality 
\begin{equation}
 0\leq \omega \leq 2K \Omega_H\ .
\label{eq:omega_hanni}
\end{equation}
We should recall that $\Omega_H$ must also satisfy Eq.~(\ref{eq:Omegamax}). 
We will search for $\Omega_H^\text{crit}$ in this region.

\subsection{WKB analysis}\label{sec:WKB}
\begin{figure}
  \leavevmode
  \begin{center}
    \begin{tabular}{ c c }
      \includegraphics[scale=0.4]{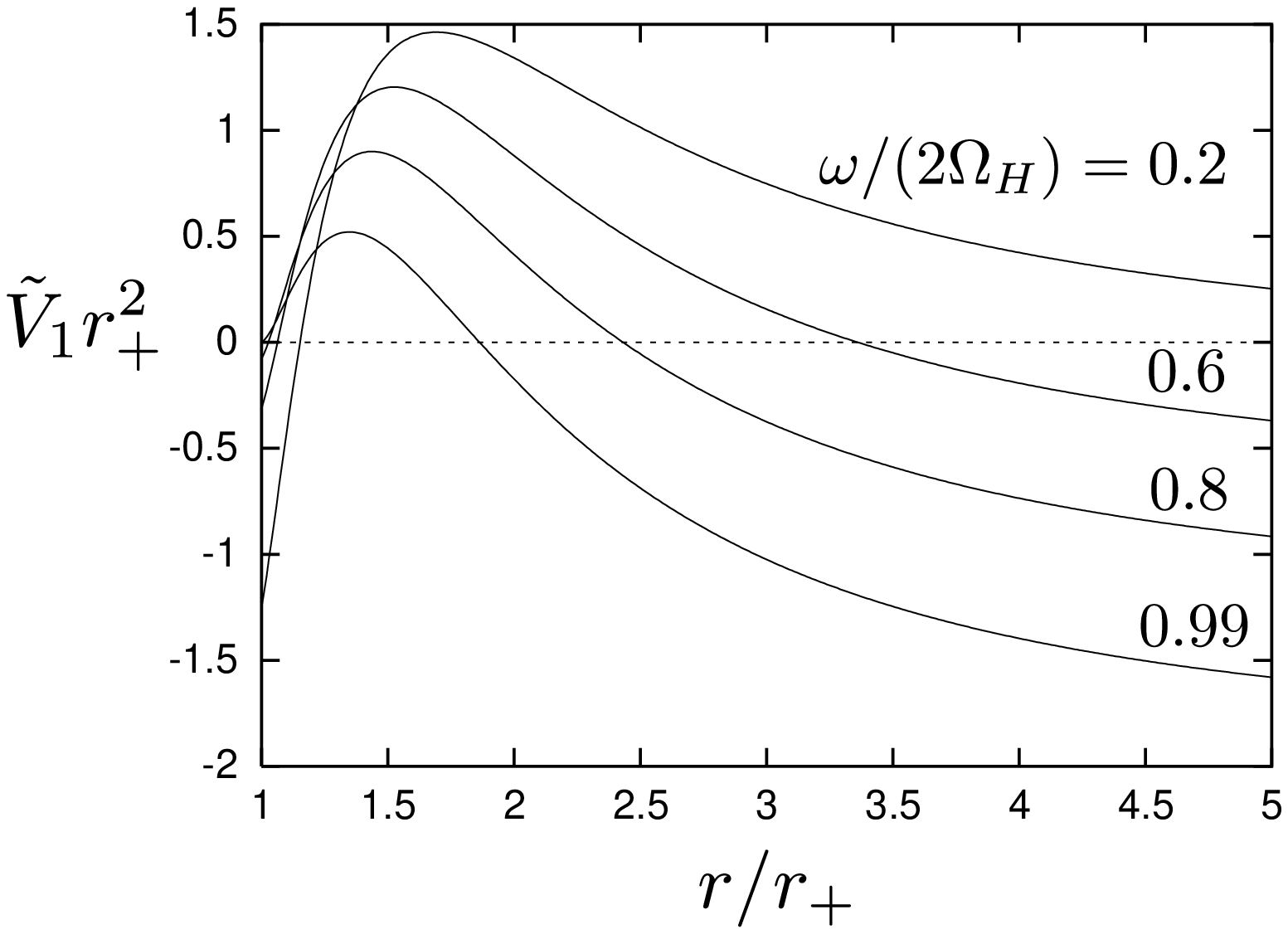}
      &
      \includegraphics[scale=0.4]{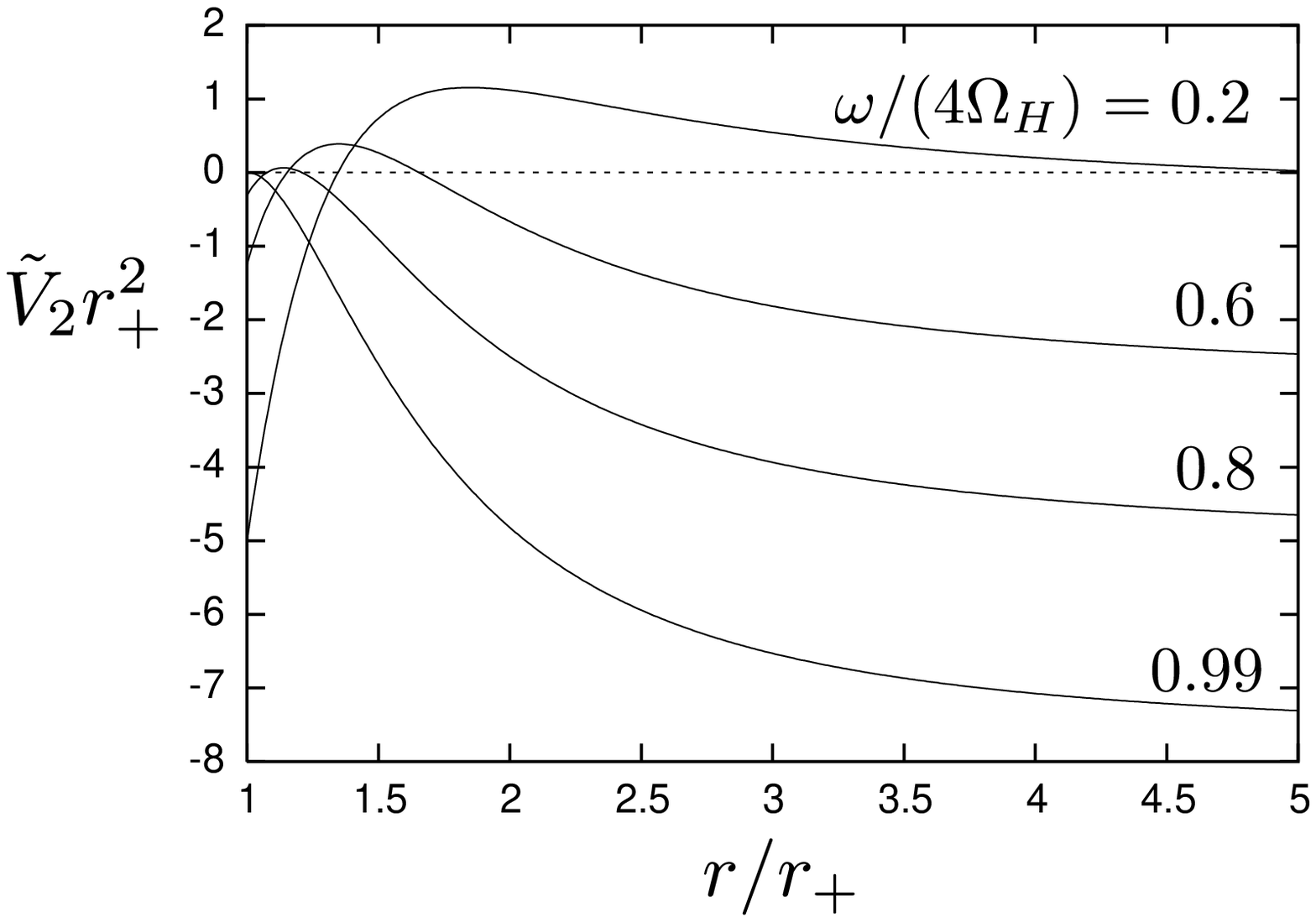}
    \end{tabular}
    \caption{$\tilde{V}_1$ and
   $\tilde{V}_2$ for $\Omega_H/\Omega_H^\text{max}=0.99$. 
   We see that $\tilde{V}_2$ tends to be negative for large
   $\omega$. However, for $\tilde{V}_1$, a positive region remains even
   for sufficiently large $\omega$.}
    \label{fig:tildeV}
  \end{center}
\end{figure}

Before the numerical analysis, we study the master
equation~(\ref{eq:master_unify})  using the WKB approximation. 
Let us define
\begin{equation}
 \tilde{V}_K( r) = V_K( r) - (\omega-2K\Omega_K( r))^2\ .
\end{equation}
If $\tilde{V}_K<0$ everywhere, one of the WKB solutions of the
master equation~(\ref{eq:master_unify}) is
\begin{equation}
 \Phi_K \sim \exp\left(i\int d r_\ast \sqrt{-\tilde{V}_K}\right)\ .
\end{equation}
The asymptotic form of this solution becomes
\begin{align}
&\Phi_K \rightarrow
 e^{-i(\omega-2K\Omega_H) r_\ast}\ ,
\quad( r\rightarrow  r_+)\nonumber\\
&\Phi_K \rightarrow e^{i\omega r_\ast}\ ,
\quad( r\rightarrow \infty)
\label{eq:WKBasym}
\end{align}
where we use Eq.~(\ref{eq:omega_hanni}). 
Equation~(\ref{eq:WKBasym}) nothing but the boundary
condition~(\ref{eq:MPbc}). On the other hand, if there is a region
satisfying $\tilde{V}_K>0$, the WKB analysis leads to
$|Z_\text{out}/Z_\text{in}|\sim 1$. Thus,  the condition
\begin{equation}
\tilde{V}( r) \lesssim 0\quad\text{for all}\quad  r\ 
\end{equation}
can be considered as a rough criterion for the existence of instability. 
Let us look at the potentials  $\tilde{V}_1$ and $\tilde{V}_2$  shown in
Fig.~\ref{fig:tildeV}. 
The potential  $\tilde{V}_2$ tends to be negative for large $\Omega_H$ and
$\omega$. For $K> 2$, 
we find a similar behavior to that of $\tilde{V}_2$. 
In the case of $\tilde{V}_1$, however, 
a positive region remains even
for sufficiently large $\omega$ and $\Omega_H$.
From the results of these WKB analyses, we can speculate that the
$(J=0,M=0,K=1)$ mode is stable
and the $(J,M,K=J+2)$ modes might be unstable for large $\Omega_H$ and $\omega$. 
With this intuition, we shall perform numerical calculations. 

\subsection{Numerical analysis}
\begin{figure}
  \leavevmode
  \begin{center}
    \begin{tabular}{ c c }
      \includegraphics[scale=0.4]{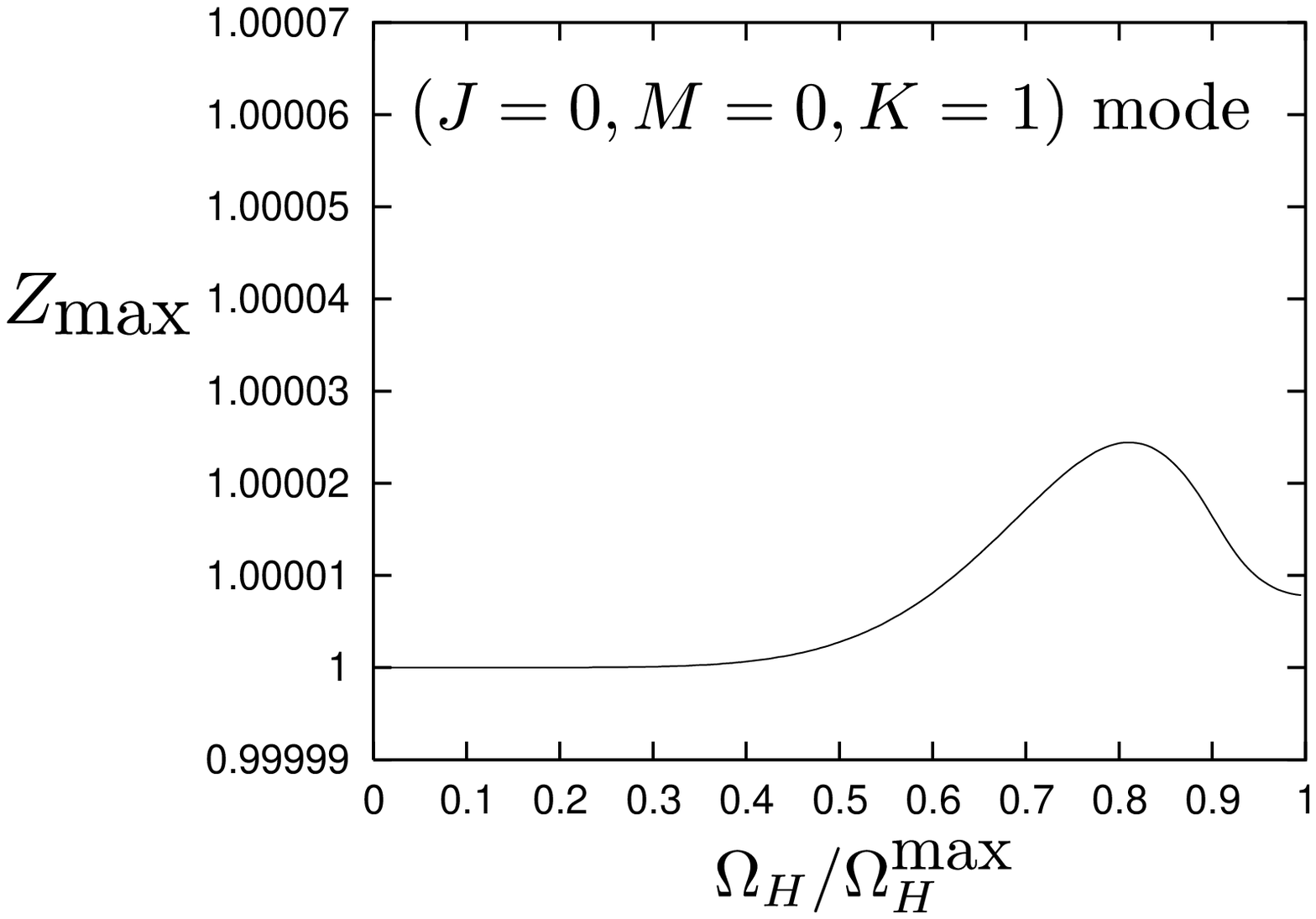}
      &
      \includegraphics[scale=0.4]{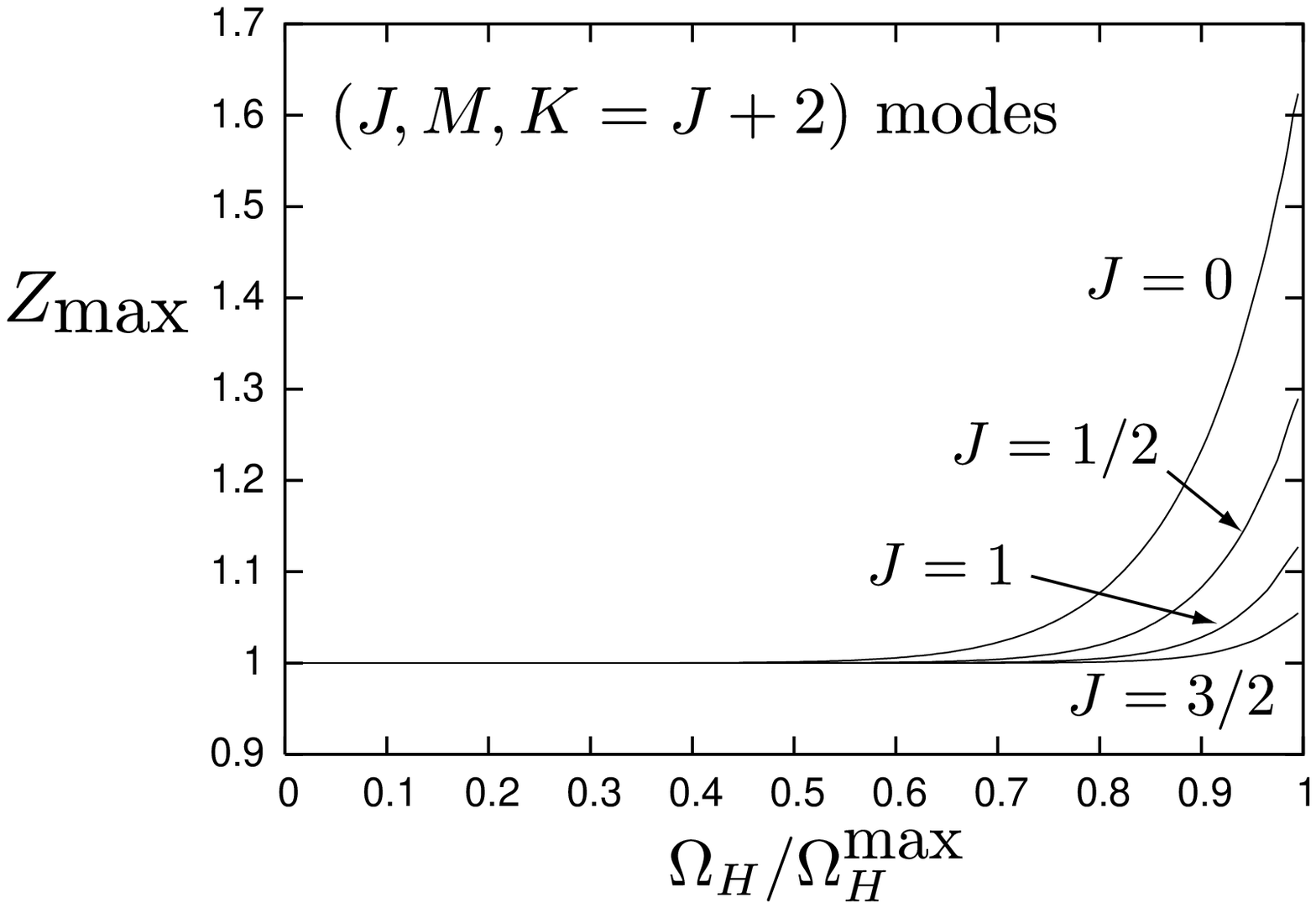}
    \end{tabular}
    \caption{Function
 $Z_\text{max}(\Omega_H)$ for $(J=0,M=0,K=1)$ mode and $(J,M,K=J+2)$ modes.  
We see that $Z_\text{max}$ is finite for $\Omega_H<\Omega_H^\text{max}$.}
    \label{fig:Zmax}
  \end{center}
\end{figure}

Now, we must solve Eq.~(\ref{eq:master_unify}) numerically. 
We integrate this master equation from 
$r_1 = (1.0+10^{-6})r_+ $ to $r_2=100.0r_+$
with the initial condition 
$\Phi_{K}=\exp(-i\omega r_{\ast})$ at $r=r_1$. This is nothing but 
the ingoing boundary condition at the horizon. At $r=r_2$, 
we checked the ratio of the amplitudes of ingoing and outgoing modes:
\begin{equation}
 Z(\omega,\Omega_H) = |Z_\text{out}|^2/|Z_\text{in}|^2\ . 
\end{equation}
We calculated this ratio  for each $\omega$ and $\Omega_H$ in domains 
  (\ref{eq:Omegamax}) and (\ref{eq:omega_hanni}). 
If $Z_\text{in}=0$ at $r=r_2$ for some $\omega$ and $\Omega_H$, the function
$Z(\omega,\Omega_H)$ would diverge.
This would be a signal of instability.  We plot
\begin{equation}
 Z_\text{max}(\Omega_H) = \max_{\text{fixed }\Omega_H} Z(\omega,\Omega_H)
\end{equation}
 in Fig.~\ref{fig:Zmax}.
For $(J=0,M=0,K=1)$,  we see $Z_\text{max}\simeq 1$.  It is
 expected from the WKB analysis in the \S\ref{sec:WKB}. 
For $(J,M,K=J+2)$, the figure shows that  $Z_\text{max}$
 takes a maximal value at $\Omega_H=\Omega_H^\text{max}$. 
This behavior is also expected  from the results of the WKB analysis in the previous subsection. 
Taking a look at Fig.~\ref{fig:Zmax}, 
at least for $\Omega_H<\Omega_H^\text{max}$,
$Z_\text{max}(\Omega_H)$ is finite.\footnote{
Note that,  in the maximally rotating case, the outer and inner horizons are
degenerate and the spacetime structure changes from that in the
$\Omega_H<\Omega_H^\text{max}$ case. Therefore, we must analyze
the case of $\Omega_H=\Omega_H^\text{max}$ separately.  }
Therefore, we  conclude that  Myers-Perry black holes are stable
 when $\Omega_H<\Omega_H^\text{max}$.  

\section{Conclusions}\label{sec:conc}
We have studied the stability of five-dimensional Myers-Perry black holes 
with equal angular momenta. Utilizing the symmetry of these solutions,
 $SU(2)\times U(1)\simeq U(2)$, we have identified the master variables 
 with which the perturbed equations can be reduced to master equations.
By analyzing the master equations, we have shown the 
 stability of Myers-Perry black holes with equal angular momenta. 
Strictly speaking, we have not shown the stability of Myers-Perry
black holes completely, because we analyzed restricted modes. 
Empirically, however, the instability appears in the lower eigenvalue modes. 
For example, the Gregory-Laflamme instability appears in the {\it s}-wave. 
Therefore, our results for $(J=0,M=0,K=0, \pm 1, \pm 2)$ modes give
strong evidence for the stability of Myers-Perry black holes.

In this paper, 
we considered the Myers-Perry black holes with equal angular momenta.
There are several qualitative arguments for the stability of general Myers-Perry black holes. 
In the case of $D\geq 6$, the argument goes as follows.
The Myers-Perry black hole with one rotating
axis reduces  to a black brane solution in the limit of
large angular momentum. Since Gregory-Laflamme instability occurs in the black brane
system~\cite{Gregory:1993vy}, 
we can speculate that the Myers-Perry black hole with one large angular
momentum is unstable in the ($D\geq 6$)-dimension~\cite{Emparan:2003sy}. 
In the case of $D=5$, we cannot use the above argument. 
Instead, we can use the fact that the black ring solution is thermodynamically more
preferable than the Myers-Perry black hole with one large angular
momentum in order to infer the instability of
the five-dimensional Myers-Perry black hole with
one large angular momentum~\cite{Emparan:2001wn}. 
Although these arguments seem to be reasonable, the direct dynamical analysis
must be performed to achieve the final conclusion.

The Myers-Perry black holes can be generalized to those with the
cosmological constant, the so-called Kerr-AdS
spacetimes~\cite{Gibbons:2004js,Gibbons:2004uw,Hawking:1998kw}.
Our method is also applicable to five-dimensional Kerr-AdS black holes
with equal angular momenta.  
It is  interesting to investigate such a case from the point of view
of the  AdS/CFT correspondence. 
From the thermodynamics of CFT$_4$, which is a dual theory of five-dimensional Kerr-AdS
spacetime,
instability can be expected for Kerr-AdS spacetime~\cite{Hawking:1998kw,Hawking:1999dp}. 
On the gravity side, there are several works on the stability of Kerr-AdS
spacetime. In the case of 4-dimensional Kerr-AdS spacetime, the superradiant 
instability has been found~\cite{Cardoso:2004hs,Cardoso:2006wa}.
In odd dimensions higher than five, 
the same instability of Kerr-AdS black holes with equal angular momenta 
has been shown to exist~\cite{Kunduri:2006qa}. 
In the case of a $D\geq 7$-dimensional Kerr-AdS black hole with one rotating axis, 
it has been shown that the superradiant instability  appears
 in the tensor-type perturbation~\cite{Kodama:2007sf}. 
However, there is no work on the stability analysis of 
five-dimensional Kerr-AdS black holes (except for a massless case~\cite{Carter:2005uw}).
It is of interest to study the stability of five-dimensional Kerr-AdS black
holes using our formalism.

\section*{Acknowledgements}
This work is supported in part by JSPS Grants-in-Aid for Scientific Research,
No. 193715 (K. M.) and 18540262 (J. S.) and 
also by the 21COE program ``Center for Diversity and
Universality in Physics'', Kyoto University. 
J. S. thanks the KITPC for their hospitality during the period when a part of
the work was carried out.

\appendix

\section{Gravitational perturbation equations for $J=M=K=0$
 mode}\label{app:eqsK=0}
In the following, we list the components of the Einstein tensor
for the $(J=0,M=0,K=0)$ mode. 
\begin{align}
\delta G_{tt}=& 
-\frac{4i\omega \mu^2 a^2}{r^6}h_{tr}'
-\frac{8i\omega \mu^2 a^2(r^2-\mu)}{r^9G(r)}h_{tr}
\notag\\
&+\frac{G(r)(3r^6-12\mu r^4+2\mu(6\mu+a^2)r^2-8\mu^2 a^2)}{2r^7}h_{rr}'
\notag\\
&+\frac{1}{r^{12}}\big[3r^{10}-6r^8\mu+2\mu(-6\mu+\mu a^2
 \omega^2-4a^2)r^6
\notag\\
&\hspace*{2cm}
+8\mu^2(3\mu+7a^2)r^4
-4a^2 \mu^2 (22\mu+3a^2)r^2+56a^4\mu^3)\big]h_{rr}
\notag\\
&-\frac{4i\omega\mu a(r^2-2\mu)}{r^6}h_{r3}'
-\frac{4i\omega\mu a(r^4-4\mu r^2+2\mu(2\mu-a^2))}{r^9G(r)}h_{r3}
\notag\\
&-\frac{2G(r)(r^2-2\mu)}{r^4}h_{+-}''
-\frac{4\mu(r^4-2(\mu + 2a^2)r^2+6\mu a^2)}{r^9}h_{+-}'
\notag\\
&+\frac{8\mu a^2((\mu\omega^2-2)r^6+2\mu
 r^4+4\mu(\mu-a^2)r^2-4a^2\mu^2)}{r^{14}G(r)}h_{+-}\ ,\\
&\notag\\
\delta G_{tr}=&
-\frac{i\omega(3{r}^4-6\mu{r}^2+2\mu{a}^2)}{2r^5}h_{rr}
-\frac{2\omega^2\mu a}{r^4G(r)}h_{r3}
\notag\\
&\hspace*{6cm}
+\frac{2i\omega}{r^2}h_{+-}'
-\frac{2i\omega(r^4+2\mu{a}^2)}{r^7G(r)}h_{+-}\ ,\\
&\notag\\
\delta G_{t3}=&
\frac{-i\omega\mu a(r^4+2\mu{a}^2)}{r^6}h_{tr}'
-\frac{2i\omega\mu a(r^4+2\mu{a}^2)(r^2-\mu)}{r^9G(r)}h_{tr}
\notag\\
&-\frac{\mu a G(r)(2r^4-3\mu{r}^2+2\mu{a}^2)}{r^7}h_{rr}'
+\frac{\mu a}{2r^{12}}\big[(\omega^2r^{10}-2r^8
\notag\\
&
+2\mu(a^2\omega^2-6)r^6+8\mu(3\mu+5a^2)r^4-88\mu^2a^2r^2+56\mu^2{a}^4)\big]h_{rr}
\notag\\
&-\frac{i\omega(r^6-2\mu{r}^4+2\mu{a}^2r^2-8{\mu}^2{a}^2)}{2r^6}h_{r3}'
+\frac{i\omega}{2G(r)r^9}[-3r^8+12\mu r^6
\notag\\
&\hspace*{2cm}
-4\mu(3\mu+a^2)r^4+24\mu^2a^2r^2
-16\mu^3a^2+4\mu^2a^4]h_{r3}
\notag\\
&+\frac{2\mu a G(r)}{r^4}h_{+-}''
+\frac{2\mu a(r^4+2\mu{r}^2-6\mu{a}^2)}{r^9}h_{+-}'
\notag\\
&+\frac{2\mu a}{r^{14}G(r)}[\omega^2r^{10}-6r^8+2\mu(a^2\omega^2+6)r^6
\notag\\
&\hspace*{4cm}
-16\mu a^2r^4+8\mu^2a^2r^2-8\mu^2a^4]h_{+-}\ ,
\end{align}
\begin{align}
\delta G_{rr}=&
-\frac{i\omega(3r^4+2\mu{a}^2)}{r^5G(r)}h_{tr}
-\frac{(3r^4-2\mu{a}^2)}{r^6}h_{rr}
+\frac{4i\omega\mu a}{r^5G(r)}h_{r3}
\notag\\
&+\frac{4(r^2-\mu)}{r^5G(r)}h_{+-}'
+\frac{2}{r^{12}G(r)^2}\big[\omega^2r^{10}-2r^8+2\mu(a^2\omega^2+4)r^6
\notag\\
&\hspace*{4cm}
-8\mu(\mu+a^2)r^4
+16\mu^2a^2r^2-8\mu^2a^4\big]h_{+-}\ ,\\
&\notag\\
\delta G_{r3}=&
\frac{i\omega \mu a}{r^3}h_{rr}
-\frac{\omega^2(r^4+2\mu{a}^2)}{2r^4G(r)}h_{r3}
-\frac{4i\omega\mu a}{r^5G(r)}h_{+-}\ ,\\
&\notag\\
\delta G_{+-}=&
-\frac{i\omega(r^4+2\mu{a}^2)}{2r^2}h_{tr}'
-\frac{i\omega(r^4-\mu{r}^2+2\mu{a}^2)(r^4-2\mu{a}^2)}{r^7G(r)}h_{tr}
\notag\\
&-\frac{G(r)(r^2-\mu)}{2r}h_{rr}'
+\frac{1}{4r^8}[\omega^2r^{10}-2r^8+2\mu(a^2\omega^2-2)r^6
\notag\\
&\hspace*{4cm}
+8\mu(\mu+a^2)r^4
-8\mu^2a^2r^2-8\mu^2a^4]h_{rr}
\notag\\
&+\frac{2i \omega \mu a}{r^2}h_{r3}'
+\frac{4i\omega\mu^2a(r^2-2a^2)}{r^7G(r)}h_{r3}
+\frac{1}{2}G(r)h_{+-}''
\notag\\
&-\frac{(r^4-6\mu{r}^2+10\mu a^2)}{2r^5}h_{+-}'
+\frac{1}{2r^{10}G(r)}[\omega^2r^{10}+4r^8
\notag\\
&
+2\mu(a^2\omega^2-8)r^6
+16\mu(\mu+2a^2)r^4-64\mu^2a^2r^2+48\mu^2a^4]h_{+-}\ ,\\
&\notag\\
\delta G_{33}=&
-\frac {i\omega(r^4+2\mu{a}^2)^2}{4r^6}h_{tr}'
-\frac{i\omega(r^2-\mu)(r^4+2\mu{a}^2)^2}{2r^9G(r)}h_{tr}
\notag\\
&-\frac{G(r)(r^8-\mu r^6+4\mu a^2r^4-6\mu^2a^2r^2+4\mu^2a^4)}{4r^7}h_{rr}'
\notag\\
&+\frac{1}{8r^{12}}\big[\omega^2r^{14}-2r^{12}+4\mu(a^2\omega^2-1)r^{10}
\notag\\
&\hspace*{1cm}
+4\mu(2\mu+5a^2)r^8
+4\mu^2 a^2(a^2\omega^2-16)r^6
\notag\\
&\hspace*{2cm}
+8\mu^2a^2(6\mu+13a^2)r^4-176\mu^3a^4r^2+112\mu^3a^6\big]h_{rr}
\notag\\
&+\frac{i\omega\mu a(r^4+2\mu{a}^2)}{r^6}h_{r3}'
+\frac{2i\omega \mu a(2r^6-3\mu r^4+4\mu{a}^2r^2-2\mu^2a^2)}
{r^9G(r)}h_{r3}
\notag\\
&+\frac{G(r)(r^4+2\mu{a}^2)}{2r^4}h_{+-}''
-\frac{r^8-6\mu r^6+8\mu a^2r^4-4\mu^2a^2r^2+12\mu^2a^4}{2r^9}h_{+-}'
\notag\\
&+\frac{1}{2r^{14}G(r)}\big[\omega^2r^{14}-2r^{12}+4\mu(a^2\omega^2-1)r^{10}
+4\mu(4\mu-3a^2)r^8
\notag\\
&\hspace*{3cm}
 +4\omega^2\mu^2a^4r^6-24\mu^2a^4r^4+16\mu^3a^4r^2
-16\mu^3a^6\big]h_{+-}\ .
\end{align}
From 
$\delta G_{tr}=\delta G_{rr}=\delta G_{r3}=\delta G_{+-}=0$, 
We can eliminate $h_{tr},h_{rr}$ and $h_{r3}$, and obtain the master
equation~(\ref{eq:K=0_master}).

\section{Derivation of master equation for $(J=0,M=0,K=1)$ mode}
\label{app:drb_master_K=1}

First, we should note there is a nondynamical 
variable $f_t$ in the action~(\ref{eq:action_K=1}).
This means there are constraints. 
To treat the constraints properly, 
it is convenient to adopt a Hamiltonian formalism~\cite{Dirac}.
From the action~(\ref{eq:action_K=1}), one can read off a Lagrangian as
\begin{align}
 \mathcal{L} =& e^A |\dot{f}_ r-f_t'|^2 + e^B|f_t|^2 + e^C|f_ r|^2 \nonumber\\
&-ie^D\{(\dot{f}_ r-f_t')f_ r^\ast - (\dot{f}_ r-f_t')^\ast f_ r\}+ie^E(f_t f_ r^\ast - f_t^\ast f_ r)\ .
\label{action}
\end{align}
Here, to simplify the calculation, we have introduced the notations
\begin{align}
&e^A = \frac{r( r^4+2\mu a^2)}{4}\ ,\quad 
e^B = \frac{4\mu^2 a^4}{ r^5 G(r)}\ , \nonumber\\
&e^C = \frac{4\mu^2a^2(r^2 -a^2)}{ r^5}\ ,\quad 
e^D = \mu a  r\ ,\quad 
e^E = 4\mu a\ .
\end{align}
The conjugate momenta of 
$ f_ r,  f_ r^\ast$ are given by
\begin{align}
  \pi_ r =& \frac{\partial \mathcal{L}}{\partial \dot{f}_ r^\ast}
  = e^A (\dot{f}_ r-f_t')+ie^D f_ r\ ,\nonumber\\
  \pi_ r^\ast =& \frac{\partial \mathcal{L}}{\partial \dot{f}_ r}
  = e^A (\dot{f}_ r-f_t')^\ast-ie^D f_ r^\ast\ .
\end{align}
The above momenta satisfy the following canonical commutation relations,
\begin{equation}
   \{f_ r(t, r),\pi_ r^\ast(t, r')\} =
  \{f_ r^\ast(t, r),\pi_ r(t, r')\} = \delta( r- r')\ ,
\end{equation}
where $\{\ ,\ \}$ represents a Poisson bracket. 
Taking the variation of the Lagrangian (\ref{action}) with respect to
$f_t^\ast$, we obtain the constraint
\begin{equation}
 f_t = e^{-B}(-\pi_ r' + ie^E f_ r)\ .
\label{eq:consraint3}
\end{equation}
The Legendre transformation gives the Hamiltonian 
\begin{align}
 \mathcal{H}_0 =&  
 \pi_ r \dot{f}_ r^\ast + \pi_ r^\ast \dot{f}_ r - \mathcal{L}\nonumber\\
 =& e^{-A}\pi_ r \pi_ r^\ast - e^B f_t f_t^\ast - (e^C-e^{2D-A}) f_ r f_ r^\ast
+ \pi_ r f_t'{}^\ast + \pi_ r^\ast f_t'\nonumber\\
&+ie^{D-A}(\pi_ r f_ r^\ast - \pi_ r^\ast f_ r) - ie^E(f_t f_ r^\ast - f_t^\ast f_ r)
\ .
\end{align}
Using the constraint equations, we obtain the physical Hamiltonian  
\begin{equation}
\begin{split}
 \mathcal{H}_\text{phys} =&\, e^{-A}\pi_ r \pi_ r^\ast +
 e^{-B}\pi_ r'\pi_ r'{}^\ast + (e^{2E-B}-e^C+e^{2D-A})f_ r f_ r^\ast \\
 &+ ie^{E-B}(\pi_ r' f_ r^\ast - \pi_ r'{}^\ast f_ r) + ie^{D-A}(\pi_ r
 f_ r^\ast - \pi_ r^\ast f_ r)  \ ,
\end{split}
\end{equation}
where total derivative terms are omitted. 
We obtain the following equation of motion for the physical variable:
\begin{eqnarray}
 \dot{f}_ r 
 &=& e^{-A}\pi_ r - (e^{-B}\pi_ r')' + i(e^{E-B}f_ r)' - ie^{D-A}f_ r\ ,  \nonumber \\
 \dot{\pi}_ r    
 &=&  -(e^{2E-B}-e^C+e^{2D-A})f_ r - ie^{E-B}\pi_ r' - ie^{D-A}\pi_ r\ .
\label{eq:Hamilton_eq1}
\end{eqnarray}
We obtain the master equation for the $(J=0,M=0,K=1)$ mode by eliminating the variable $f_ r$
from Eq.~(\ref{eq:Hamilton_eq1}).


\end{document}